\documentclass[preprint,12pt,a4paper]{elsarticle}

\makeatletter
\def\ps@pprintTitle{%
  \let\@oddhead\@empty
  \let\@evenhead\@empty
  \def\@oddfoot{\reset@font\hfil\thepage\hfil}
  \let\@evenfoot\@oddfoot
}
\makeatother

\usepackage{amssymb}

\usepackage{cancel}

\usepackage{hyperref}

\usepackage[fleqn]{amsmath}
\usepackage[normalem]{ulem}

\newcommand{\msout}[1]{\text{\sout{\ensuremath{#1}}}}

\begin{document}

\begin{frontmatter}




\title{The chimera revisited:\\Wall- and magnetically-bounded turbulent flows}


\author[au1]{Nils T. Basse}
\ead{nils.basse@ri.se}

\address[au1]{RISE Research Institutes of Sweden \\ Brinellgatan 4, 504 62 Borås, Sweden \\ \vspace{10 mm} \small {\rm \today}}

\begin{abstract}
This paper is a first attempt at bringing together various concepts from research on wall- and magnetically-bounded turbulent flows. Brief reviews of both fields are provided: The main similarities identified are coherent (turbulent) structures, flow generation and transport barriers. Examples are provided and discussed.
\end{abstract}


\end{frontmatter}



\section{Introduction}

\subsection{Turbulent flows in fluids and plasmas}

Turbulent flow in fluids, i.e. liquids and gases, has been studied since long before da Vinci's contributions \cite{marusic_a}; usage of the term "turbulence" can be traced back to da Vinci \cite{colagrossi_a}, although consistent usage of the nomenclature did not occur until the early twentieth century \cite{schmitt_a}. In contrast, plasmas, i.e. ionised gases, were discovered much later and named by Langmuir in 1928 \cite{langmuir_a}. Over the following decades it was found that turbulence is as important in plasmas as in fluids, see \cite{kadomtsev_a} and references therein. Plasmas can also be described as fluids and their behaviour in electromagnetic (EM) fields is termed magnetohydrodynamics (MHD) \cite{goedbloed_a}. Broader efforts to treat fluids and plasmas together exist \cite{NatAcad,choudhuri_a,pop_pof}, but direct comparisons are scarce. Common phenomena in the turbulent flows of fluids and plasmas have previously been identified, such as momentum transport caused by velocity fluctuations. With this paper we have the ambition to provide more details on known similarities, but also to discover new shared mechanisms and by that provide impetus for novel research directions: We aim to facilitate cross-pollination between turbulent flow research in fluids and plasmas.

As representatives of fluids and plasmas we choose two types of turbulent flows:
\begin{itemize}
\item Fluids: Wall-bounded \cite{pope_a}
\item Plasmas: Magnetically-bounded \cite{horton_a}
\end{itemize}

Probably wall- and magnetically-bounded turbulent flows have not been systematically associated for several reasons, e.g. (i) because the plasma physics (PP) community first and foremost focuses on links with astrophysical plasmas and (ii) the fluid mechanics (FM) community has not been considering parallel efforts in the PP community in a systematic fashion. However, FM research, for example the Kolmogorov 1941 (K41) energy cascade \cite{kolmogorov_a,frisch_a}, has been used for PP turbulence studies. Other concrete examples are covered in e.g. \cite{terry_a, diamond_a}.

This paper is personal in the sense that it is a result of my own "voyage through turbulence" \cite{davidson_a} from PP to FM research \footnote{ \samepage I began in academic PP research (1997-2005), where I also had contact with K41 and energy/enstrophy cascades for two- and three-dimensional flows. Moving to industrial FM research (2006-2023), my focus was on wall-bounded turbulent flows, e.g. turbulent mixing of gases, two-phase flow, flow noise (acoustics) and thermofluids. This paper is an attempt to synthesise my experience, but there is a risk of not referring to the latest research; particularly for PP, since I have not been active in the field since 2005. Wall- and magnetically-bounded flows are treated, but cross-disciplinary efforts have also included un-bounded flows, e.g. similarities between turbulence on mm and Mpc scales \cite{basse_a,basse_b}. However, those results are outside the scope of the present paper.}

\subsection{Motivation behind the paper}

The transport barrier (TB) concept was introduced to FM by Prandtl as the laminar/turbulent boundary layer (LBL/TBL) \cite{prandtl_a,schlichting_a,davidson_a}, where the LBL constitutes an edge transport barrier (ETB) using nomenclature from magnetic confinement fusion PP. The boundary layer (BL) is characterised by mean velocity shear and molecular (LBL)/turbulent (TBL) viscosity. The ETB was first identified in magnetically confined fusion plasmas in 1982 \cite{wagner_a} and named the high (H) confinement mode as opposed to the previously known low (L) confinement mode.

In 1995, uniform momentum zones (UMZs), regions where the streamwise momentum is close to being constant, were discovered \cite{meinhart_a}. The UMZs are separated by internal shear layers. That year, internal transport barriers (ITBs) were discovered in two magnetically confined fusion experiments \cite{levinton_a,strait_a}.

In this paper, we will attempt to link the LBL and the ETB concepts and the UMZ and ITB concepts - for the first time to the best of our knowledge. In addition to the LBL/ETB and UMZ/ITB similarities, other observations which triggered this paper include:

\begin{itemize}
  \item Increasing core fluctuations for the pipe flow high Reynolds number ($Re$)\footnote{ \samepage We use $Re$ without subscript as a general term; later in the paper, two specific definitions, the bulk and friction Reynolds number, are defined using subscripts.} transition \cite{basse_d} is similar to controlled confinement transitions in fusion plasmas \cite{zoletnik_a, basse_c}
  \item Travelling wave solutions in pipe flow \cite{wedin_a} are reminiscent of the magnetic field structure (islands) in fusion plasmas
\end{itemize}

\noindent \hrulefill \\

In order to avoid copyright issues, figures from cited papers will be discussed but not shown. This unfortunately makes the paper more difficult to read, but open source versions of most references can be found online.

The paper is organised as follows: Sections \ref{sec:wall} and \ref{sec:magnetic} consist of primers (in the spirit of The Los Alamos Primer \cite{serber_a}) on wall- and magnetically-bounded turbulent flows, respectively. In Section \ref{sec:TB} transport barriers are treated in general and Section \ref{sec:core} focuses on comparing turbulent flows in the core region. An overview of important concept similarities and differences follows in Section \ref{sec:overview}. A discussion can be found in Section \ref{sec:disc} and we conclude in Section \ref{sec:conc}.

\section{Wall-bounded turbulent flows}
\label{sec:wall}

In FM, there are two main ways to treat wall-bounded turbulent flows; one is the statistical approach and the other is a dynamical systems viewpoint \cite{dennis_a}. An important difference is that the (traditional) statistical approach considers turbulent flows with high $Re$, whereas the dynamical systems analysis is limited to lower $Re$. We will focus on the statistical point of view below but will discuss the dynamical systems approach in Section \ref{subsec:FM_dynamical}. Research on the laminar-turbulent pipe flow transition \cite{avila_a} identifies a third perspective, which is linear or nonlinear hydrodynamic stability. This has been deemed out of scope for this paper and will not be covered.

Canonical, i.e. standard, wall-bounded flows include zero pressure gradient (ZPG) TBLs, channels and pipes \cite{smits_a}. In the following, we focus on pipe flow but will also address features of other canonical flows.

The coordinates are usually named as (i) streamwise ($x$ along the flow), (ii) wall-normal ($y$ perpendicular to the wall) and (iii) spanwise ($z$ parallel to the wall and perpendicular to the streamwise direction).

We assume the no-slip and no-penetration boundary conditions (BCs) \cite{mckeon_a}, i.e. that the velocity at the wall is zero and that the walls are impermeable.

\subsection{Transition from laminar to turbulent flow}
\label{subsec:FM_lam_turb}

To define the bulk Reynolds number $Re_D$, where $D=2R$ is the pipe diameter and $R$ is the pipe radius, the area-averaged streamwise mean flow velocity $U_m$ is used:

\begin{equation}
Re_D = \frac{D U_m}{\nu_{\rm kin}},
\end{equation}

\noindent where $\nu_{\rm kin}$ is the kinematic molecular viscosity.

At a certain $Re_D$ ($\sim 2 000$), the laminar to turbulent transition takes place \cite{reynolds_a,davidson_a}, associated with a steepening of the edge velocity gradient. However, the transition is gradual with $Re_D$; as it increases, what is observed first are turbulent puffs, which are regions of turbulence separated by laminar regions. Turbulent puffs either decay or split, both with very long timescales. As $Re_D$ increases further, the turbulent patches increase in size and become what is called slugs, before turbulent flow fills the entire pipe \cite{avila_a}.

\subsection{The boundary layer concept}
\label{subsec:FM_BL}

Both velocity (momentum) and temperature (heat) BLs exist in wall-bounded laminar and turbulent flows \cite{schlichting_a}. The concepts are analogous, with a region of (velocity/temperature) gradients close to the wall and another region of (almost) constant values towards the pipe axis. The thermal BL can either be coupled to the velocity field or not  depending on the conditions, e.g. assumptions on density, dynamic viscosity, specific heat capacity and thermal conductivity.  

\subsection{The turbulent/non-turbulent interface}
\label{subsec:FM_TNTI}

In addition to a BL close to the wall, TBLs also have a turbulent/non-turbulent interface (TNTI) at the free-stream boundary where the TBL ends \cite{corrsin_a, ishihara_a, eisma_a}.

The TNTI was identified in \cite{corrsin_a} from experiments and characterised as a thin fluid layer where viscous forces dominate, the "laminar superlayer", thought to be a wrinkled sheet of viscous vortical fluid. The mean and fluctuating vorticity propagate through this (wrinkled) layer to the nonturbulent (irrotational) region. The thickness of the layer is found to be of the order of the Kolmogorov length:

\begin{equation}
\eta_K = \left( \frac{\nu_{\rm kin}^3}{\varepsilon} \right)^{1/4},
\end{equation}

\noindent where $\varepsilon$ is the dissipation rate of $k$, the turbulent kinetic energy (TKE) per unit mass.

Direct numerical simulations (DNS) of TBLs were presented in \cite{ishihara_a}, where a small peak in the spanwise vorticity and an associated small jump in streamwise velocity was observed at the TNTI. The interfacial layer was found to have an inertia-viscous double structure:

\begin{itemize}
\item A turbulent sublayer, with a thickness $l_I$ (between interface and vorticity peak) of the order of the Taylor microscale:
\begin{equation}
\lambda_T \approx \sqrt{10 \nu_{\rm kin}\frac{k}{\varepsilon}}
\end{equation}
\item An outer boundary (superlayer), thickness $l_S$ (width of vorticity peak) of order of the Kolmogorov length scale $\eta_K$
\end{itemize}

The length scale of the turbulent sublayer $l_I$ is longer than the length scale of the outer boundary $l_S$.

Analysis shows that the TNTI acts as a barrier in both directions: Exterior irrotational fluctuations are being damped/filtered at the interface and internal rotational fluctuations are also blocked at the TNTI which remains sharp.

Velocity jumps at the TNTI and inside the TBL were studied experimentally in \cite{eisma_a} and found to have similar characteristics. The velocity jump height was found to be constant for $y/\delta_{99}>0.5$, i.e. far from the wall, with larger jumps closer to the wall. Here, $\delta_{99}$ is the (99\%) TBL width, where $\delta$ corresponds to $R$ in a pipe. The internal layers are regions of high shear which are thought to bound large scale motions (LSM), see Section \ref{subsec:FM_turb_struct}. The jump thickness $\delta_w$ is observed to scale with the (local) Taylor microscale: $\delta_w \approx 0.4 \lambda_T$. The internal layers are observed to move away from the wall, with a faster layer velocity further from the wall. It is conjectured that shear layers are generated not only at the wall, but away from the wall as well.

\subsection{Mean turbulent flow}
\label{subsec:FM_mean}

The mean flow is in the streamwise direction, with three main wall-normal regions: the viscous sublayer closest to the wall, the logarithmic (log) layer and the wake region towards the pipe axis \cite{coles_a, pope_a}. Sometimes the terms inner (outer) layer are used for the regions close to (far away) from the wall, respectively. 

\subsection{Fluctuating turbulent flow}
\label{subsec:FM_fluc}

Streamwise velocity fluctuations have a peak close to the wall (the inner peak) and a second peak in the log region which becomes more prominent with increasing $Re$ (the outer peak). The inner peak has fixed wall-normal position (normalised to the viscous length scale), but it is under discussion if it has a maximum or continues to increase with $Re$. The attached eddy model (AEM) \cite{townsend_a,marusic_b} leads to structures increasing in size from the wall towards the pipe axis, also as a (streamwise and spanwise, but not wall-normal) log-law, but decreasing towards the pipe axis as opposed to the mean streamwise flow \cite{marusic_c}.

Streamwise velocity fluctuations are usually higher than both the wall-normal and spanwise fluctuations; energy transfer takes places from the streamwise to the wall-normal and spanwise fluctuations \cite{schlichting_a}.

\subsection{Turbulence models}
\label{subsec:FM_turb_models}

Turbulence models attempt to close the equations of motion, e.g. by introducing a turbulent (eddy) viscosity; for the simplest algebraic model, the turbulent viscosity is proportional to the mixing length, which is a concept introduced by Prandtl \cite{prandtl_b,prandtl_c}. The turbulent shear (streamwise/wall-normal) Reynolds stress (RS) $\tau_{xy}$ is then equal to the product of the dynamic turbulent viscosity $\mu_t$ and the mean velocity gradient $\mathcal{S}=|\partial U/\partial y|$:

\begin{equation}
\label{eq:RS}
\tau_{xy} = \mu_t \times \left| \frac{\partial U}{\partial y} \right| = n_f \nu_t \times \mathcal{S},
\end{equation}

\noindent where $n_f$ is the fluid density and $\nu_t$ is the kinematic turbulent viscosity: $\nu_t \gg \nu_{\rm kin}$. The turbulent RS represents the turbulent transport of momentum to the wall due to velocity fluctuations.

\subsection{Turbulent structures}
\label{subsec:FM_turb_struct}

Turbulence consists of smaller structures in the inner layer, whereas both small and large structures coexist in the outer layer. The structures can be sorted into four different groups \cite{smits_b}:
\begin{itemize}
  \item Sublayer (near-wall) streaks generated by streamwise vortices \cite{avila_a}
  \item Hairpin or $\Lambda$ vortices
  \item Vortex packets or LSM
  \item Even larger structures, called (i) very large scale motions (VLSM) in pipe flow and (ii) superstructures in boundary layers
\end{itemize}

The hairpin or $\Lambda$ vortices are vorticity structures with a "head" and two "feet"; the head is typically further downstream than the feet, i.e. the vortices are leaning in the streamwise direction.

There is an ongoing discussion on the interaction between structures - whether large structures in the outer layer are superimposed onto inner layer structures or if the mechanism is amplitude modulation \cite{marusic_d,andreolli_a}. There is also a discussion whether the large structures are "active" or "passive", i.e. whether they contribute to the turbulent shear RS or not \cite{deshpande_a}.

An area of research that has traditionally been included in the statistical approach but which also contains elements of the dynamical systems viewpoint is proper orthogonal decomposition \cite{berkooz_a}, which has e.g. been used to analyse radial and azimuthal modes of VLSM \cite{smits_b}.

\subsection{Minimal flow unit}
\label{subsec:FM_MFU}

A minimal flow unit (MFU) has been identified \cite{jimenez_a} which is a minimum structure size needed to sustain small-scale turbulence close to the wall. This has been done using DNS to isolate small structures in the inner layer.

The spanwise MFU $\lambda_z^+ = \lambda_z u_{\tau}/\nu_{\rm kin} \approx 100$, where "+" indicates normalisation by the viscous length scale $\nu_{\rm kin}/u_{\tau}$. Here, $u_{\tau}$ is the friction velocity. The spanwise MFU matches the value widely observed for the spacing of sublayer streaks and streamwise vortices. The streamwise MFU was observed to be $\lambda_x^+ \approx 250-350$, which is of the same order as experimental observations of vortices near a wall. Turbulence statistics are in good agreement with simulations covering the entire cross-section below a wall-normal distance $y^+ = 40$; near-wall turbulence can be sustained indefinitely for a layer width of this size.

Subsequent work on MFUs \cite{hwang_a} found two different streamwise MFUs:
\begin{itemize}
\item $\lambda_x^+ \simeq 200-300$: Quasi-streamwise vortices
\item $\lambda_x^+ \simeq 600-700$: Near-wall streaks
\end{itemize}

\subsection{Turbulent length scales}
\label{subsec:FM_turb_len}

We have already introduced the Kolmogorov and Taylor length scales in Section \ref{subsec:FM_TNTI}. Two other useful scales can be added, the first being the mixing length (mentioned in Section \ref{subsec:FM_turb_models}):

\begin{equation}
\ell_m = \sqrt{\frac{\nu_t}{\mathcal{S}}},
\end{equation}

\noindent and the second being the length scale of larger eddies:

\begin{equation}
L = \frac{k^{3/2}}{\varepsilon},
\end{equation}

\noindent see \cite{basse_d} and the associated Supplementary Information for more details.

The Kolmogorov scale is the smallest scale and $L$ is the largest scale. The Taylor and mixing length scales are intermediate (meso), with the Taylor length being shorter than the mixing length.

For the log-law region we can write:

\begin{equation}
\mathcal{S}=|\partial U/\partial y|=\frac{u_{\tau}}{\ell_m},
\end{equation}

\noindent and define a length scale associate with the mean velocity gradient:

\begin{equation}
L_U = U/\mathcal{S} = \frac{U}{|\partial U/\partial y|},
\end{equation}

\noindent which can be used to rewrite the mean velocity gradient as:

\begin{equation}
\label{eq:ML}
\mathcal{S}=U/L_U=\frac{u_{\tau}}{\ell_m}
\end{equation}

Two other length scales have also previously been mentioned, the largest (outer) scale $\delta$ (or $R$) in Section \ref{subsec:FM_TNTI} and the small (inner) viscous length scale in Section \ref{subsec:FM_MFU}. The ratio between these scales defines the friction Reynolds number:

\begin{equation}
Re_{\tau} = \frac{\delta u_{\tau}}{\nu_{\rm kin}} = \frac{u_{\tau}}{2 U_m} Re_D
\end{equation}

From these length scales, it has been argued that mixed scaling can be relevant, i.e. combinations of the inner and outer length scales, for example \cite{mckeon_a}:

\begin{equation}
y_m = \sqrt{\frac{y}{\delta} y^+} = \frac{y^+}{\sqrt{Re_{\tau}}} = \frac{y}{\delta} \sqrt{Re_{\tau}}
\end{equation}

\subsection{Uniform momentum zones}
\label{subsec:FM_UMZ}

The first type of internal TBL observed was the UMZ, with nearly constant streamwise momentum separated by thin viscous-inertial shear layers \cite{meinhart_a}. In the shear layers, spanwise vorticity is lumped into strongly vortical regions, i.e. a collection of vortices. This interpretation differs from the picture in \cite{corrsin_a}, where the TNTI was interpreted as a continuous vortex sheet.

Later observations in TBLs have continued to study the UMZ structure and the intense vorticity in the shear layers \cite{deSilva_a}. The number of UMZs increases proportionally to $\log(Re_{\tau})$ and the UMZ thickness increases with increasing distance from the wall. The structures generating the UMZ behave consistently with the AEM: Hairpin packets are shown to create a zonal-like organisation.

An UMZ vortical fissure (VF) model was presented in \cite{cuevasBautista_a} and validated against DNS simulations of channel flow. The UMZs are segregated by narrow fissures of concentrated vorticity, with a discrete number of fissures (internal shear layers) across the TBL. The model has two primary domains, (i) an inertial domain and (ii) a subinertial domain; the theoretical basis for the inertial layer (far from wall) is more solid than for the subinertial layer (near-wall). A fixed fissure width gives the best match to DNS and the jump in streamwise velocity is proportional to $u_{\tau}$. The wake is not taken into account for the modelled mean velocity. The internal VFs are allowed to be repositioned (from an initial master profile) and a momentum-exchange mechanism is necessary:

\begin{itemize}
\item Outward flux of vorticity is connected with inward flux of momentum
\item The VF characteristic velocity is recalculated:                        

{\begin{itemize} \item If the VF moves farther from (toward) the wall, there is momentum loss (gain) compared to the master profile \end{itemize}}

\item The outermost VF is not allowed to move and exchange momentum                     
\end{itemize}

The momentum-exchange mechanism, i.e. that VFs gain (lose) momentum when they are displaced toward (away) from the wall, is consistent with a variation of the streamwise/wall-normal turbulent RS:

\begin{equation}
\frac{1}{n_f} \frac{{\rm d} \tau_{xy}}{{\rm d} y} = \overline{v \omega_z} - \cancelto{0}{\overline{w \omega_y}},
\end{equation}

\noindent where overbar is time averaging and $\omega_z$ is the spanwise vorticity. The last right-hand side term is zero because only wall-normal VF movement is considered.

An alternative concept to the UMZ model, a momentum transport barrier (MTB) model, has been published in \cite{aksamit_a}. 

\subsection{Quiescent core}
\label{subsec:FM_quiescent}

For turbulent channel flow, what is known as the quiescent core has been experimentally identified and characterised \cite{kwon_a}. The quiescent core is a large UMZ, which can cover up to 40-45\% of the channel; it can be approximated by regions where the mean velocity is above 95 \% of the centerline (CL) mean velocity: $U>0.95U_{CL}$. The interface has a jump in streamwise velocity, and sometimes - but not always - a vorticity peak. Inside the core UMZ, the streamwise velocity varies only weakly. The core UMZ is meandering (moves around), can reach the wall and be streamwise separated (breakup). The core UMZ has low TKE, i.e. it is weakly turbulent (quiescent).

A two-state model of the TBL (extendable to internal flows) is presented in \cite{krug_a} to capture the log-law and law of the wake regions. The new model has a log-law state and a free stream state, with a velocity jump at their interface. The concept for mean flow can be applied to streamwise turbulence as well. One drawback of the model is that it does not take the viscous region close to the wall into account. The model is calibrated against measurements and the position of the interface is fitted to a Gaussian distribution which is independent of $Re_{\tau}$. The resulting velocity jumps and deviations of the fit from the log-law are also independent of $Re_{\tau}$ except for pipe flow below $Re_{\tau}=3 400$, which is interesting and may be related to the high $Re$ transition region for pipe flow \cite{basse_d}.

Open channel flow was studied in \cite{pirozzoli_a} using DNS, and it was concluded that: "The virtual absence of a wake region and of corrective terms to the log-law in the present flow leads us to conclude that deviations from the log-law observed in internal flows are likely due to the effects of the opposing walls, rather than the presence of a driving pressure gradient." Thus, the law of the wake may only exist due to TBL interactions.

\subsection{Uniform thermal zones}
\label{subsec:FM_UTZ}

After the identification of UMZ, uniform thermal zones (UTZ) have been found, which consist of regions of relatively uniform temperature separated by thermal interface layers \cite{yao_a}. The analysis was done on DNS simulations of transcritical channel flow. An analogy was made between UMZ and momentum internal interface layers (MIILs) and UTZ and thermal internal interface layers (TIILs). Thus, the two types of zones relate to velocity (momentum) and temperature (heat) fields. A local heat transfer peak is expected in the TIILs. The MIILs and TIILs were found to be at similar but not identical locations, i.e. not collocated.

A model of UTZ and TIILs has been published in \cite{ebadi_a}, constructed along the same lines as the UMZ model in \cite{cuevasBautista_a}. The nomenclature is slightly modified compared to \cite{yao_a}; here, the uniform thermal zones are called uniform temperature zones and the TIILs are named thermal fissures (TF). The heat model (UTZ/TF) is combined with the momentum model (UMZ/VF) and calibrated against DNS simulations of channel flow. As for the momentum model, the TFs can move (from an original master profile) and exchange heat as they move in the wall-normal direction: If a TF moves towards (away from) the wall, its temperature increases (decreases), respectively. The finding in \cite{yao_a} that the VF/TF (MIIL/TIIL) are correlated but not coincident is confirmed in \cite{ebadi_a}.

It is important to note that temperature is a passive scalar \cite{warhaft_a} (when buoyancy is neglected), i.e. it does not affect the dynamics of the fluid.

\subsection{Uniform concentration zones}
\label{subsec:FM_UCZ}

Experiments have identified a third type of uniform zone (UZ), uniform concentration zones (UCZ) \cite{eisma_b}: As is the case for temperature, concentration is also a passive scalar.

In both shear and shear-free flows, ramp-cliff (RC) structures have been identified for passive scalars, i.e. a slow increase (ramp) followed by a fast decrease (cliff) \cite{sreenivasan_a}. These structures have also been said to have a "saw-tooth appearance" with plateaus separated by cliffs \cite{shraiman_a}. From an interpretation of experiments, the RC structures can be understood as large counter-rotating structures which form a saddle point associated with converging-diverging separatrices as discussed in \cite{antonia_a}. The cliff (or front) occurs at the diverging separatrix, which has an inclination close to the direction of the principal axis of strain. If the passive scalar is temperature, the front is the separation between warm and cold fluids entrained in the counterflowing structures. In aircraft measurements, inverse cliff-ramp (CR) structures have been considered signatures of the Kelvin-Helmholtz instability \cite{wroblewski_a}.

\subsection{Uniform momentum and temperature zones}
\label{subsec:FM_UMZ_UTZ}

Simultaneous existence of both UMZ and UTZ has been reported for both stably and unstably stratified turbulent flow by analysis of large eddy simulations (LES) \cite{heisel_a, salesky_a}. In \cite{heisel_a} the stably stratified planetary boundary layer (PBL) was treated; it was found that UMZ and UTZ are "closely, but not perfectly related". Unstably stratified channel flow was covered in \cite{salesky_a}, where it was found that: "Conditional averaging indicates that both UMZ and UTZ interfaces are associated with ejections of momentum and warm updrafts below the interface and sweeps of momentum and cool downdrafts above the interface."

\subsection{Turbulence control}
\label{subsec:FM_control}

Methods for classical flow control up to around the year 2000 have been covered in \cite{gad-el-hak_a}. Methods can be active or passive, e.g.:

\begin{enumerate}
\item Passive: Riblets, surface treatment, tripping, shaping
\item Active: Suction, blowing, wall cooling/heating
\end{enumerate}

Here, the purpose can be e.g. to modify transition to turbulence, to decrease friction (pressure drop), to enhance heat transfer and to reduce acoustic noise \cite{gad-el-hak_b}.

More recent work includes turbulence suppression due to pulsatile driving of pipe flow \cite{scarselli_a}. The work was inspired by the human cardiovascular system, where blood flow in the aorta is an example of pulsating flow. By comparing experiments and DNS, it is demonstrated that both turbulence and turbulent drag can be reduced significantly in pulsating flow.

Machine learning (ML) has in recent years become a more powerful tool for both turbulence simulation and control \cite{vinuesa_a}. The method can be seen as a fourth pillar complementing theory, experiments and simulations.

\subsection{Dynamical systems viewpoint}
\label{subsec:FM_dynamical}

For the dynamical systems approach, we focus on invariant solutions to the Navier-Stokes equations (NSE) as defined in \cite{budanur_a}:

"Here by ‘invariant solutions’ or ‘exact coherent structures’ we mean compact, time-invariant solutions that are set-wise invariant under the time evolution and the continuous symmetries of the dynamics. Invariant solutions include, for instance, equilibria, travelling waves, periodic
orbits and invariant tori. Note in particular that the closure of a relative periodic orbit is an invariant torus."

The first exact coherent state (ECS) or travelling wave (TW) solution to the NSE was identified theoretically in \cite{nagata_a} followed by multiple efforts, both with theoretical \cite{waleffe_a, faisst_a, wedin_a, schneider_a, pringle_a} and experimental \cite{hof_a, dennis_a, jaeckel_a} focus.

For pipe flow, it has been found that the ECS originate in saddle-node bifurcations at $Re_D$ down to around 400 \cite{paranjape_a}. The TWs consist of a certain number of azimuthally and radially separated streaks, for example threefold azimuthal symmetry: 6 outer (high speed) streaks and 3 inner (low speed) streaks. The TWs lead to transport of slow fluid towards the center and transport of fast fluid towards the wall.

Additional TW solutions were constructed in \cite{wedin_a} by "mixing three key flow structures - 2-dimensional streamwise rolls, streaks and 3-dimensional streamwise-dependent waves - in the right way". This is in line with what has been termed the self-sustaining process (SSP), see \cite{waleffe_b} and references therein. Here, it is proposed that edge turbulence is maintained (against viscosity) by a cycle of rolls, streaks and waves. 

Another process has been proposed for core turbulence \cite{montemuro_a}, which involves inertial ECS, in contrast to the viscous ECS for the SSP . It is interesting to notice the appearance of "Kelvin's cat's-eyes vortex pattern" inside the VF, see Figures 9 and 12 in \cite{montemuro_a}.

A main obstacle to a direct link between the dynamical and the statistical approach is to identify invariant solutions for high $Re$. Experimental support that these solutions exist have come from \cite{dennis_a}, where ECS are shown to have an impact up to $Re_D = 35 000$. 

Other theoretical ECS solutions have been investigated in parallel, we refer to related work focusing on the relative periodic orbit (RPO) framework \cite{cvitanovic_a, cvitanovic_b, budanur_a, cvitanovic_c}. The two types of ECS solutions can be summarised as:

\begin{itemize}
\item TW: A fixed velocity profile moving in the streamwise direction with a constant phase speed
\item RPO: Time-dependent velocity profiles which repeat exactly after a certain time period and streamwise length; in addition, these orbits may also have azimuthal rotations
\end{itemize}

A dynamical systems approach has also been pursued in studies of the laminar-turbulent transition \cite{avila_a}; as mentioned, TWs have been identified for $Re_D$ lower than the observed transition. It has also been shown that spatially localised RPOs can experience a series of bifurcations leading to transient chaos.

\section{Magnetically-bounded turbulent flow}
\label{sec:magnetic}

For the material on PP, we focus on commonalities with FM, therefore many specific features have been disregarded. Of course this entails a risk of leaving out important topics. An example of what is left out is specific issues relating to EM fields and plasma currents.

A note on units: In PP, temperature is usually stated using units of energy, where 1 eV corresponds to around 11 600 K. Another convention to keep in mind is that for PP, density has the units of particle density (number of particles per volume), whereas in FM, mass density is used (mass per volume).

\subsection{Magnetic field structure}
\label{subsec:PP_magfield}

A plasma consists of charged particles (electrons and ions), which need to be confined within a toroidal shape to enable fusion. Since charged particles follow magnetic field lines (with superimposed gyroradii), the method of confinement is to construct closed magnetic field surfaces.

The basic shape of a magnetic confinement devices is a torus, with coordinates (i) toroidal (the "long" way around a torus), (ii) radial and (iii) poloidal (the "short" way around a torus).

Relating to pipe flow, the corresponding coordinates are toroidal/streamwise, radial/wall-normal and poloidal/spanwise.

Additional (a) perpendicular and (b) parallel coordinates refer to the directions perpendicular (cross-field) and parallel to the magnetic field. These are different from - but related to - the toroidal, radial and poloidal coordinates.

We focus on cases from tokamaks \cite{wesson_a} but include material on stellarators and heliotrons \cite{wakatani_a} when relevant.

For these machine types, the main toroidal magnetic field is generated by external planar coils. A main difference between tokamaks and stellarators/heliotrons is how the poloidal magnetic field is created: In tokamaks, it is created by a toroidal current induced through transformer action, but in stellarators/heliotrons it is created by modular (non-planar) coils. For stellarators, the modular coils are predominantly poloidal whereas for heliotrons, the modular coils are mainly toroidal. This implies that the plasma current in tokamaks is much higher than in stellarators/heliotrons, which has important implications for e.g. current-driven instabilities, steady-state operation and machine complexity.

All machine types treated herein generate an MHD equilibrium with nested magnetic surfaces. The boundary is named the last closed flux surface (LCFS) which is called a separatrix if it includes one or more "X-points", which are points with zero (null) poloidal field. Plasmas can also be bound by physical limiters. We use the term "magnetically-bounded" for plasmas which are bounded by a separatrix, i.e. where the LCFS is not in contact with physical surfaces. The region between the separatrix and the physical wall is called the scrape-off layer (SOL), where magnetic field lines are open and intersect the wall. Divertors intersect the open field lines from the separatrix and are used for particle and heat exhaust.

The winding number of the magnetic field lines is called the safety factor in tokamaks due to its importance for plasma stability:

\begin{equation}
q = \frac{{\rm d} \phi}{{\rm d} \theta},
\end{equation}

\noindent where $\phi$ is the toroidal angle and $\theta$ is the poloidal angle. Traditionally, another definition has been used in stellarators/heliotrons:

\begin{equation}
\msout{\iota} = \frac{\iota}{2\pi} = \frac{1}{q}
\end{equation}

Typically, $q$-profiles in tokamaks have a minimum $q_{\rm min}$ close to the axis and increase towards the plasma edge. For stellarators/heliotrons, the $\msout{\iota}$-profile is often more flat. We define the magnetic shear:

\begin{equation}
s = \frac{r}{q} \frac{\partial q}{\partial r},
\end{equation}

\noindent where $r$ is the minor radius measured from the magnetic axis.

Thus, tokamaks have high shear and stellarators/heliotrons have low shear.

The magnetic field decreases from the center of the torus outwards inversely proportional to the major radius $R$, which can lead to particle trapping due to the magnetic mirror effect. For tokamaks, these are called banana orbits and centered on the outboard midplane (the low field side). For stellarators/heliotrons, the particles are helically trapped.

\subsection{Turbulence and improved confinement regimes}
\label{subsec:PP_confinement}

As mentioned, the purpose of the magnetic field is confinement; the plasma also needs to have a sufficiently high temperature for the ions to fuse and release energy. Two timescales can be used to quantify energy and particle confinement, namely the energy confinement time $\tau_E$ and the particle confinement time $\tau_p$. These timescales indicate how efficient the confinement of energy (temperature) and particles (density) is.

Another way of gauging confinement quality is $\beta$, which is the plasma pressure normalised to the magnetic pressure. $\beta$ can be defined both using the total ($B$), the toroidal ($B_{\phi}$) or the poloidal ($B_{\theta}$) magnetic field. As the plasma pressure increases, the center of the magnetic axis is displaced radially outwards, an effect called the Shafranov shift.

If transport is only taking place due to thermal motion (Coulomb collisions), with curvature effects included, it is called neoclassical transport \cite{hinton_a}. However, in reality much larger transport is observed perpendicular to the magnetic field, which is called anomalous transport \cite{carreras_a}.

Anomalous transport is caused by turbulence, e.g. microinstabilities driven by the ion (ITG) or electron (ETG) temperature gradient or by trapped electrons such as the trapped electron mode (TEM). The smallest turbulent scale is due to ETG, medium scale due to TEM and largest scale due to ITG. Instabilities driven by density or temperature gradients are called drift waves (DW). Even larger scale (macroscopic) MHD instabilities can be driven by e.g. current, pressure or fast particles. Often instabilities can be ballooning, which means that - due to curvature effects - their growth rate is larger on the outer side of the torus compared to the inner side. Turbulence can lead to the formation of streamers, first identified in nonlinear gyrokinetic simulations of ETG turbulence \cite{jenko_a,dorland_a} followed by theoretical predictions for ITG turbulence \cite{diamond_b}. Streamers are radially elongated mesoscale vortices centered on the outboard midplane; they lead to enhanced cross-field transport, thereby degrading confinement.

A main effort in the fusion community is to understand and reduce anomalous transport to improve confinement and obtain more efficient fusion reactions.

One way to control anomalous transport is by external heating of electrons and ions, for example by ion or electron cyclotron resonance heating (ICRH/ECRH) or by neutral beam injection (NBI). The plasma current can also be manipulated both using external heating and current drive, e.g. lower hybrid current drive (LHCD).

The plasma state can experience either gradual confinement improvements or sudden bifurcations to improved confinement regimes; sometimes improved confinement is associated with instabilities such as edge localised modes (ELMs), which lead to bursts of cross-field transport of particles and energy. Other improved confinement regimes can be associated with coherent modes which regulate transport and avoid ELMs.

\subsection{Length scales}
\label{subsec:PP_turb_len}

An important group of length scales is associated with the Larmor radius, which is the gyration distance of charged particles around the magnetic field:

\begin{equation}
\rho_j = \frac{m_j v_{\perp}}{e_jB} = \frac{v_{\perp}}{\omega_{cj}},
\end{equation}

\noindent where the subscript $j$ represents electrons ($e$) or ions ($i$), $m_j$ is the mass, $v_{\perp}$ is the velocity perpendicular to the magnetic field, $e_j$ is the charge and $\omega_{cj}=e_j B/m_j$ is the cyclotron frequency. Here, we can relate the velocity to temperature by assuming two degrees of freedom:

\begin{equation}
v_{\perp}^2 = 2 v_{Tj}^2,
\end{equation}

\noindent which leads to:

\begin{equation}
\rho_j = \sqrt{2} \frac{m_j v_{Tj}}{e_j B} = \sqrt{2} \frac{v_{Tj}}{\omega_{cj}}
\end{equation}

For scaling purposes, the ion Larmor radius normalised to the minor radius of the machine ($r=a$) is used:

\begin{equation}
\rho^* = \frac{\rho_i}{a},
\end{equation}

\noindent and for turbulence modelling, the ion Larmor radius at the electron temperature is used:

\begin{equation}
\rho_s = \sqrt{2} \frac{m_i v_{Te}}{e_i B}
\end{equation}

Scale lengths have been mentioned previously in Section \ref{subsec:FM_turb_len}; we generalise the notation to write the scale length $L_x$ of a quantity $x$ as:

\begin{equation}
L_x = \frac{x}{|{\rm d} x/{\rm d} r|} = (|{\rm d} (\ln x)/{\rm d} r|)^{-1}
\end{equation}

Equation (\ref{eq:ML}) can be reformulated for electron density fluctuations ($x=n_e$):

\begin{equation}
\frac{n_e}{L_{ne}} = \frac{\delta n_e}{\rho_s},
\end{equation}

\noindent where $\delta n_e$ are density fluctuations (corresponding to the friction velocity) and $\rho_s$ is the typical scale of the density fluctuations (corresponding to the mixing length). For DWs, the density fluctuations saturate at this level:

\begin{equation}
\frac{\delta n_e}{n_e} = \frac{\rho_s}{L_{ne}} \sim \frac{1}{k_{\perp} L_{ne}},
\end{equation}

\noindent where $k_{\perp} \sim 1/\rho_s$ is the perpendicular wavenumber of the density fluctuations.

Microscales are on the order of the (ion/electron) Larmor radius, from sub-mm to mm scales, depending on temperature and magnetic field strength. Macroscales are on the order of the machine minor radius and mesoscales are between micro- and macroscales; an example of a mesoscale phenomenon is streamers, and we will encounter other mesoscale structures later.

An effect known as turbulence spreading, originally theoretically predicted in \cite{garbet_a}, occurs for inhomogeneous turbulence \cite{singh_a}: "Turbulence spreading is a process of turbulence self-scattering by which locally excited turbulence spreads from the place of excitation to other places." This is not related to the K41 paradigm which deals with homogeneous turbulence.

\subsection{Rational safety factors and transport}
\label{subsec:PP_safety}

If $q = m/n$ is a rational number ($m$ and $n$ both integers), then the magnetic field line returns to the initial position after $m$ toroidal and $n$ poloidal rotations. For a fixed toroidal angle, this corresponds to a poloidal mode number $m$ and for a fixed poloidal angle it corresponds to a toroidal mode number $n$.

Since the magnetic field line paths constitute a Hamiltonian system, rational values of the safety factor correspond to resonant tori, which are unstable against perturbations according to the Kolmogorov-Arnold-Moser (KAM) theorem \cite{ott_a}. Perturbations can lead to the formation of magnetic islands or ergodic regions.

A classical example of instabilities is sawtooth crashes (relaxations) for $q < 1$, where heat and particles are ejected from the core plasma due to magnetic reconnection \cite{zweibel_a}: "Magnetic reconnection is a topological rearrangement of magnetic field
that converts magnetic energy to plasma energy." The periodic core temperature collapse is due to an instability which has an $m=n=1$ structure, corresponding to $q=1$.

Enhanced transport has been observed for $q$-profiles at or close to low-order rationals in the Rijnhuizen Tokamak Project (RTP) \cite{lopesCardozo_a, hogeweij_a}. Transport barriers for the electron temperature were observed as temperature steps which could be controlled by the deposition location of external electron heating. A "q-comb" model was constructed to model the transport barriers as low electron heat conductivity at low-order rationals, possibly due to the formation of magnetic island chains.

As for the RTP tokamak, a similar behaviour has been observed in the Wendelstein 7-Advanced Stellarator (W7-AS) \cite{brakel_a, brakel_b}. Here, reduced transport was also found to be associated with low-order rationals.

\subsection{Magnetic islands caused by instabilities or topology}
\label{subsec:PP_islands}

In both tokamaks and stellarators/heliotrons, magnetic islands can be caused by instabilities as mentioned above, e.g. global Alfv\'en eigenmodes (GAE) \cite{weller_a} and tearing modes \cite{fitzpatrick_a}. These islands can be either non-rotating ("locked") or rotating.

In addition, natural magnetic islands can exist in stellarators/heliotrons. An example is from the W7-AS and Wendelstein 7-X (W7-X) stellarators, where islands form for

\begin{equation}
\msout{\iota} = \frac{n}{m} = \frac{5}{m},
\end{equation}

\noindent the constant "5" being due to the fact that the machines have a five-fold toroidal symmetry. The five field periods are also flip symmetric, leading to ten identical sections. For W7-AS the standard divertor configuration (SDC) was $m=9$ \cite{mccormick_a} whereas for W7-X it is $m=5$ \cite{feng_a}, the change being due to $\msout{\iota}$-profile differences. Thus, W7-X has larger islands with lower poloidal mode numbers compared to W7-AS.

The natural magnetic islands can be used to form a separatrix and an associated island divertor. This also enables detachment, which is a state where a large fraction of the power is dissipated by volume radiation before it reaches the physical wall. This is a potential exhaust solution under reactor conditions, since the heat flow will be intercepted before reaching the divertor target plates, leading to significantly reduced fluxes at the targets.

\subsection[]{$E \times B$ flow shear decorrelation}
\label{subsec:PP_EcrossB}

A mechanism to reduce turbulent transport by velocity shear has been theoretically identified in \cite{biglari_a}\footnote{ \samepage Earlier theoretical efforts can be found in e.g. \cite{lehnert_a}.} and reviewed along with experimental evidence in \cite{terry_a}. It causes eddy stretching which leads to eddies losing coherence (breakup), i.e. energy transfer from large scales (low wavenumbers) to small scales (high wavenumbers). It is called sheared $E \times B$ flow and is generated by the radial electric field $E_r$ which results from the radial force balance (ignoring the RS term):

\begin{equation}
E_r = \frac{1}{n_i Z_i e} \frac{{\rm d}p_i}{{\rm d}r} + v_{\phi i} B_{\theta} - v_{\theta i}B_{\phi},
\end{equation}

\noindent where the "$i$" subscript refers to ions (dominating compared to electrons), $p$ is the pressure, $Z$ is the charge state, $e$ is the electronic charge, $v_{\phi}$ is the toroidal velocity and $v_{\theta}$ is the poloidal velocity. Suppression of turbulence takes place if the shearing rate $\omega_{E \times B}$ is larger than the maximum linear growth rate $\gamma_{\rm max}$ of the relevant instability:

\begin{equation}
\omega_{E \times B} = \frac{R B_{\theta}}{B_{\phi}} \frac{\partial}{\partial {\rm r}} \left( \frac{E_r}{R B_{\theta}} \right) > \gamma_{\rm max}
\end{equation}

The shearing rate increases with shear in the radial electric field $\partial E_r/\partial r$, so the regions where the radial electric field changes rapidly as a function of radius are the regions where turbulence is suppressed most efficiently. $E \times B$ shearing is a mean flow effect on turbulence which affects not only the turbulence amplitude, but also the "phase angle between an advected fluctuation and the advecting flow" \cite{terry_a}. Shear suppression is a universal, self-regulating process between shear flow and transport: Turbulence reduction leads to steepened gradients (temperature, density), which increases the pressure gradient, which in turn increases the flow shear and reduces turbulence further.

In addition to the shearing rate criterion, three additional requirements have to be fulfilled:

\begin{itemize}
  \item The shear flow must be stable
  \item Turbulence must remain in the flow shear region for longer than an eddy turnover time \cite{pope_a}
  \item Dynamics should be 2D
\end{itemize}

These requirements are often met in fusion plasmas, but rarely in nonionised fluids; some exceptions are mentioned in \cite{terry_a}, e.g. stratospheric geostrophic flow and perhaps the laminar phase between bursts of turbulence for wall-bounded flows.

\subsection{Transport barriers}
\label{subsec:PP_TB}

In this section we provide a brief overview of the different TB variants in fusion plasmas: (i) ETB \cite{wagner_a, gohil_a}, (ii) ITB \cite{levinton_a, strait_a, wolf_a, ida_a} and (iii) both ETB and ITB \cite{gohil_b}.

\subsubsection{ETB}

As mentioned in the Introduction, the first ETB was identified in 1982 in the Axially Symmetric Divertor Experiment (ASDEX) tokamak \cite{wagner_a}. For NBI power above a certain threshold, an L-H-mode transition was obtained. This was possible for diverted plasmas but not for limited plasmas. Apart from the power threshold, H-mode could only be accessed for a safety factor at the edge $q_a>2.6$.

The improved H-mode confinement was seen as an increased poloidal $\beta$ ($\beta_p$) and an increase of the electron density and temperature. Bursts of $H_{\alpha}-D_{\alpha}$ emission were observed in H-mode which were later identified as ELM signatures.

The H-mode ETB is quite robust and has steep density and temperature gradients just inside the LCFS. $E \times B$ flow shear is part of the prerequisite for the H-mode, along with suitable edge plasma conditions which may vary between different machine designs. As of now, there is no comprehensive, predictive theory-based model for ETB formation and spatial structure.

ELMs generated by the large pressure gradients created in ETBs can often degrade or even destroy the barrier. Some methods exist to stabilise instabilities, for example applying an external magnetic field or operating variants of H-modes with quasi-coherent (QC) or edge harmonic oscillations (EHO), which provide increased particle transport through barrier withhout significantly increasing the energy transport.

\subsubsection{ITB}

As referred to in the Introduction, the first ITBs were identified in 1995 in two tokamaks, the Tokamak Fusion Test Reactor (TFTR) \cite{levinton_a} and the Doublet III-D (DIII-D) \cite{strait_a}.

For both machines, the most important component to achieve an ITB was to get reversed magnetic shear which was obtained by creating a hollow current density profile. This was done by a combination of current ramping and NBI and took advantage of the fact that the current diffusion time is much longer than the rise time of the plasma current.

The ITB led to reduced particle and ion thermal transport in the plasma core where reversed shear was created. The high pressure gradient generated strong off-axis bootstrap current which helped to maintain the hollow current density profile. Electron thermal transport was also reduced but not as significantly as the ion thermal transport. The ion thermal diffusivity and electron particle diffusivity decrease to close to or below the neoclassical level.

MHD modes can exist outside the ITB and act to limit the obtainable $\beta$.

ITBs in tokamaks were reviewed in \cite{wolf_a}. It was found that low or reversed magnetic shear in combination with large $E \times B$ shear flows are essential ITB ingredients, where magnetic shear stabilises high-$n$ ballooning modes and $E \times B$ shear stabilises medium- to long-wavelength turbulence, i.e. ion thermal transport and particle transport. It is possible to have high electron thermal transport even with ITBs. The $q$ value at 95\% of the magnetic flux, $q_{95}$, is found to be important for magnetic stability and $q_{\rm min}$ has been seen to correlate with the ITB foot. The Shafranov shift can have a stabilising effect on turbulence called $\alpha$-stabilisation. ITBs can exist with equal ion ($T_i$) and electron ($T_e$) temperatures, but also for cases where $T_i<T_e$ or $T_e<T_i$, depending on the plasma density and external heating method.

In the rest of the section we summarise results from the most recent review \cite{ida_a} which covers both tokamak and helical (in our case: Stellarators/heliotrons) plasmas. A systematic approach is applied, with an ITB definition being a (radial) discontinuity of temperature, flow velocity or density gradient.

ITBs are characterised by three parameters:

\begin{enumerate}
  \item Normalised temperature gradient $R/L_T = R \times |\nabla T|/T$ (large value: weak, small value: strong)
  \item Location $r_{\rm ITB}/a = \rho_{\rm ITB} = (\rho_{\rm shoulder}+\rho_{\rm foot})/2$ (large value: large, small value: small)
  \item Width $W/a = \rho_{\rm foot}-\rho_{\rm shoulder}$ (large value: wide, small value: narrow)
\end{enumerate}

Here, $L_T = T/|\nabla T|$ is the temperature scale length, "shoulder" is at the top of the steep gradient and "foot" is at the bottom of the steep gradient.

The key elements for ITB formation are summarised as:

\begin{itemize}
  \item Radial electric field shear ($E \times B$ flow shear)
  \item Magnetic shear
  \item Rational surface and/or magnetic islands
\end{itemize}

It is instructive to write the equations relating radial fluxes (particle, momentum, electron/ion heat) and gradients (density, toroidal rotation, temperature). For the particle flux $\Gamma$ we write:

\begin{equation}
\label{eq:particle}
\frac{\Gamma(r)}{n_e} = - \left[ \frac{D \nabla n_e}{n_e} -v_{\rm conv} \right],
\end{equation}

\noindent where $D$ is the diffusion coefficient, $n_e$ is the electron density and $v_{\rm conv}$ is the convection velocity. For the momentum flux $P_{\phi}$ we write:

\begin{equation}
\label{eq:momentum}
\frac{P_{\phi}(r)}{m_i n_e} = -\nu_{\perp} \nabla v_{\phi} + v_{\rm pinch} v_{\phi} + \Gamma_{\phi}^{\rm resi},
\end{equation}

\noindent where $m_i$ is the ion mass, $\nu_{\perp}$ is the perpendicular kinematic viscosity, $v_{\rm pinch}$ is the momentum pinch velocity and $\Gamma_{\phi}^{\rm resi}$ is the radial flux due to residual stress \cite{zhao_b,rice_a}. We note that Equation (\ref{eq:momentum}) - when disregarding the two final right-hand side terms - has the same structure as Equation (\ref{eq:RS}). For the electron and ion heat flux ($Q_{e,i}$) we write:

\begin{equation}
\label{eq:heat}
\frac{Q_{e,i}(r)}{n_e} = - \frac{\chi_{e,i} \nabla T_{e,i}}{m_i},
\end{equation}

\noindent with the electron and ion thermal diffusivity $\chi_{e,i}$. To cite \cite{ida_a}: "When the density, velocity, and temperature gradient become large due to the decrease in the diffusion coefficient, $D$, viscosity, $\mu_{\perp}$, and thermal diffusivity, $\chi_{e,i}$, the region in the plasma is called the transport barrier." We will use "diffusion coefficient" as a collective term for $D$, $\mu_{\perp}$ and $\chi_{e,i}$. An ITB can be defined as a bifurcation in the flux-gradient relationship, which causes a discontinuity in the density/velocity/temperature gradient for a given particle/momentum/heat flux, leading to the formation of a discontinuity in the gradient with radius.

The ITB foot (point) often follows integer $q$ values, typically $q=1$ ($\rho \sim 0.3$), $q=2$ ($\rho \sim 0.5$) and $q=3$ ($\rho \sim 0.7$); this is valid for positive or weakly reversed magnetic shear, but not strong reversed magnetic shear. Sometimes ITBs are also observed for half-integer $q$ values. For reversed magnetic shear, an ITB appears when $q_{\rm min}$ crosses a rational surface.

Experiments using resonant magnetic perturbations (RMPs) to produce magnetic islands were carried out in the Large Helical Device (LHD) to distinguish the role of magnetic islands and rational surfaces. It was found that \cite{ida_a}: "This experiment supports the idea that the magnetic island at the rational surface contributes to the transition from the L-mode to the ITB rather than to the rational surface itself." A reduction of transport inside magnetic islands has been observed, close to what is called the "O-point", as opposed to the previously mentioned X-point. There is a reduction in turbulence (and transport) at the boundary of magnetic islands and the pressure profile is flat in the O-point inside the islands.

Important observed differences between tokamak and helical plasmas include:

\begin{itemize}
  \item Ion barriers are most significant for tokamaks, electron barriers for helical devices
  \item Simultaneous ion/electron barriers have been seen in tokamaks, but not in helical devices
  \item In general, magnetic shear is negative for helical devices, but both positive and negative for tokamaks
  \item Differences in particle transport: Clear density barrier for tokamaks, barrier disappears for higher density in helical devices. But it exists for both when pellet injection is used.
  \item The toroidal angular velocity is higher for tokamaks
  \item The sign of the impurity pinch is opposite: Inwards for tokamaks (impurity accumulation), outwards for helical systems
   \item ITBs are more variable for tokamaks due to the freedom of the current profile (magnetic shear), which is restricted in helical devices
  \item Radial electric field:
        \begin{enumerate}
          \item Helical: Mainly generated by poloidal velocity
          \item Tokamak: Significant contribution from toroidal rotation
        \end{enumerate}
\end{itemize}

Non-locality of ITB plasmas has been observed, e.g. coupling between the inside and the outside of the ITB. The curvature of the ion temperature ($\partial^2 T_i / \partial r^2$) has been linked with ITB stability, where a convex (concave) curvature means a less (more) stable ITB, respectively.

\subsubsection{Both ETB and ITB}

It was already demonstrated in \cite{strait_a} that an ITB can coexist with both L- and H-mode edges, where an ITB with an H-mode edge is a double barrier (DB), i.e. an ETB and an ITB. Non-locality has also been observed for this type of DB, where the ITB formation takes place simultaneously with the L-H transition \cite{ida_a}.

Multiple barriers have been reviewed in \cite{gohil_b} and we present a summary of this work in the rest of the section.

The leading mechanisms for stabilisation are stated as (i) $E \times B$ flow shear and (ii) reduction of growth rates due to $\alpha$-stabilisation. The combination of ETB and ITB is useful if it can:

\begin{itemize}
\item Increase the plasma volume with reduced transport
\item Lead to improved stability against MHD modes
\item For tokamaks: Improve the bootstrap current fraction for steady-state operation
\end{itemize}

On the other hand, potential drawbacks include:

\begin{itemize}
\item ITB degradation due to the ETB, e.g. reduction of rotation shear and pressure gradient at the ITB location
\item High density at the ETB can reduce NBI penetration efficiency 
\item ELMs can lead to flattening of ITB temperature gradients
\end{itemize}

An example where the barriers lead to additive beneficial effects is the quiescent double barrier (QDB) mode in DIII-D, where an ITB is combined with a quiescent H-mode (QH) which has an EHO.

\subsection{Zonal flows}
\label{subsec:PP_ZF}

We now review zonal flows (ZFs) based on the material in \cite{diamond_a, goncalves_a, itoh_a, fujisawa_a, ida_a, zhao_a, nishizawa_a, conway_a}.

ZFs are azimuthally symmetric band-like $E \times B$ shear flows with mode numbers $n=m=0$. They are mesoscale electric field fluctuations with zero mean frequency and finite radial wave number $k_r$. ZFs are flows which are driven by turbulence, e.g. turbulent shear RS \cite{diamond_c} or DW. Due to their structure, ZFs are benign repositories for free energy and do not drive radial (energy or particle) transport. ZFs vary rapidly in the radial direction. For toroidal plasmas having a strong toroidal magnetic field (valid assumption in this paper), ZFs are predominately poloidally directed with velocities $v_{\theta} = -E_r/B$ and $v_{\phi} = -2q v_{\theta} \cos \theta$. The convention is that $\theta=0^{\circ}$ at the outboard midplane and increases in the counterclockwise direction.

ZFs differ from mean $E \times B$ shear flows (see Section \ref{subsec:PP_EcrossB}); mean shear flows are generated as a result of the ion radial force balance and ZF shear flows are driven by turbulence. Mean shear flows can persist without turbulence, whereas ZF shear flows cannot. This is reflected in the different radial electric fields:
\begin{itemize}
  \item The radial electric field from ZFs is oscillatory, complex, consists of small structures and is driven exclusively by nonlinear wave interaction processes.
  \item The mean radial electric field evolves on transport timescales and is driven by e.g. heating, fuelling and momentum input which determine equilibrium profiles, in turn regulating the radial force balance.
\end{itemize}
The mean and ZF shear flows can interact, e.g. mean flows can suppress ZFs through turbulence decorrelation. Both flow types can tilt and break turbulent eddies.

ZFs shear or quench turbulence to extract energy from it leading to a self-regulating mechanism with a predator-prey system of turbulent energy (prey) and ZF energy (predator). In that sense, ZFs can shift (delay) the onset of turbulence, often referred to as the "Dimits shift" \cite{dimits_a}.

ZFs have been linked to rational $\msout{\iota}$ values, e.g. in the H-1 National Facility (H-1NF), which was a 3-field period heliac. ZFs were found at two locations (due to reversed shear) where $\msout{\iota}=7/5$.

Because of the 3D nature of shear flow physics, several RS terms can contribute to ZF generation, e.g. radial-parallel, radial-perpendicular and radial-poloidal.

ZFs are not Landau (wave) damped but mainly collisionally damped due to friction between trapped and circulating ions; they increase with decreasing collisionality.

The energy partition between ZFs and turbulence is key for plasma confinement: A large fraction of ZFs results in better confinement. To understand the process, one can write the ratio of ZFs and turbulence as:

\begin{equation}
\zeta = V^2/N = \frac{\gamma_L/\alpha}{\gamma_{\rm damp}/\alpha} = \gamma_L/\gamma_{\rm damp},
\end{equation}

\noindent where $V$ is the ZF intensity, $N$ is the turbulence energy, $\gamma_L$ is the DW (turbulence) linear growth rate, $\alpha$ is a coupling constant between ZFs and DWs and $\gamma_{\rm damp}$ is the flow damping of ZFs due to collisionality. The ratio $\zeta$ increases with improved confinement since the damping rate decreases.

ZFs may take the role of a trigger for confinement transitions, possibly at the L-H transition. An interaction between mean and zonal flows may also exist; e.g that the mean $E \times B$ flow exists before the transition and that the additional effect of ZFs triggers the transition itself.

In nature, the Jovian belts/zones and the terrestrial jet stream have been given as examples of ZFs.

Finally we mention zonal fields which is the generation of structured magnetic fields from turbulence, i.e. a magnetic counterpart to ZFs. They were theoretically predicted in \cite{gruzinov_a} and experimentally detected in \cite{fujisawa_b}. The magnetic field structures, also with $n=m=0$ and finite radial wave number, can be generated by DW turbulence and may have a back-reaction on turbulence via magnetic shearing.

\subsection{Geodesic acoustic modes}
\label{subsec:PP_GAM}

We proceed with a review of geodesic acoustic modes (GAMs) based on material in \cite{itoh_a, fujisawa_a, zhao_a, conway_a}.

In many respects, GAMs are similar to ZFs: GAMs also have mode numbers $n=m=0$, but couple to pressure/density fluctuations with $m=\pm 1$ (poloidal mode number) and $n=0$. These fluctuations are poloidally asymmetric and highest at the top and bottom of tokamak plasmas. For stellarators/heliotrons, the highest fluctuation is not at the top and bottom, but follows the helical pitch. For completeness, we note that there is also a magnetic component with $m=\pm 2$ and $n=0$. GAMs have velocities $v_{\theta} = -E_r/B$ and $v_{\phi} = q^{-1} v_{\theta} \cos \theta$.

GAMs have a finite frequency as opposed to ZFs which have zero frequency. The GAM frequency scales with the square root of the temperature; this can be derived from single-fluid ideal (all dissipative processes neglected) MHD \cite{freidberg_a}:

\begin{equation}
\omega_{\rm GAM}^2 = \frac{2c_s^2}{R^2} \left( 1+ \frac{1}{2q^2} \right),
\end{equation}

\noindent where:

\begin{equation}
c_s = \sqrt{\gamma (T_e+T_i)/m_i}
\end{equation}

\noindent is the speed of sound and $\gamma=5/3$ is the specific heat ratio.

GAMs are both Landau damped ($\propto \exp(-q^2)$) and collisionally damped; the zero frequency ZFs are not Landau damped, but only collisionally damped. Due to the differences in Landau damping and magnetic configuration in tokamaks and helical devices, GAMs are mainly found at the edge of tokamaks and in the low $\msout{\iota}$ core region of stellarators/heliotrons. Generally, it has also been observed that GAMs are stronger (and have been observed more often) in tokamaks than helical devices.

GAMs can be driven directly from the poloidally symmetric $m=0$ component of the turbulent shear RS, similar to ZFs: "Since both the GAM and the ZF are driven by turbulence there is the issue of competition in the nonlinear transfer leading to the dominance of one or other mode." \cite{conway_a}. However, ZFs and GAMs can coexist and transitions between ZFs and GAMs have also been observed.

Both ZFs and GAMs have comparable radial correlation lengths, which are mesoscale as found for streamers as well.

The response of ZFs and GAMs to fluctuations is different: ZFs are incompressible (slow response) and GAMs are compressible (fast response).

Usually, GAMs are not observed in H-mode.

The impact of GAMs on transport can be summarised as:

\begin{itemize}
\item No direct radial energy or particle transport
\item Oscillatory flow shearing
\item Act as an energy sink through Landau damping or dissipation
\item Modulate cross-field transport through pressure fluctuations (GAMs are rarely contiguous and stable)
\end{itemize}

Finally, we collect quotes from \cite{conway_a} on the relationship between GAMs and magnetic islands:

\begin{itemize}
\item "The interaction of GAMs with MHD modes (static and rotating)
is multi-fold. An island chain may create a GAM-like oscillation, or it may enhance and/or entrain a natural edge GAM, or it may suppress and destroy the natural GAM."
\item "At the extreme, the velocity shearing associated with the GAM can also restrict the island radial structure and thus limit the growth of the MHD mode."
\item "The flow and turbulence behaviour can be divided into three distinct spatial regions: inside the island separatrix, around the island boundary, and spatially (radially) well away from the island chain."
\end{itemize}

\subsection{Blobs}
\label{subsec:PP_blobs}

Blobs are filaments generated by edge plasma turbulence with enhanced levels of particles and heat aligned along magnetic field lines in the SOL \cite{garcia_a}. There is intermittent eruptions of plasma and heat into the SOL which leads to radial motion of blobs. They are ballooning, with more transport at the outboard midplane. The fluctuation level and turbulence-driven transport (number of events) increases with $\beta$ and collisionality. Blobs have an  asymmetric waveform with time, where the rise time is fast and the decay is slow; their total duration is of order 25 ms. 

A theory on blob creation based on breakup of streamers due to velocity shear has been experimentally validated in \cite{bisai_a}. These streamers are located outside the separatrix, so in that sense they are different from the streamers previously mentioned. A possible mechanism for the shear flow generation is the interchange instability, which is "very similar in nature to the Rayleigh-Taylor instability in fluid dynamics" \cite{wakatani_a}. More blobs are observed in L-mode than in H-mode.

\section{Transport barriers}
\label{sec:TB}

\subsection{General}

For FM TBs, the edge BLs or core internal interface layers (IILs) both result from the momentum balance of the NSE - and the energy equation for thermal barriers.  For FM, the question is if there is a phenomenon analogous to the magnetic field in PP - perhaps ECS?

TBs for PP are associated with both mean $E \times B$ shear turbulence suppression and properties of the confining magnetic field, e.g. magnetic islands.

In general, both particle and heat TBs contribute to an increased pressure gradient, whereas the momentum TB results in a steeper velocity gradient.

\subsection{Edge}
\label{subsec:edge}

The LBL concept from FM seems to be equivalent to the ETB of PP associated with the H-mode. For FM, the steeper velocity gradient for turbulent flow is associated with the domination of inertial forces over viscous forces quantified by $Re$. For PP, the H-mode is associated both with an external power threshold, magnetic field effects and mean $E \times B$ shear flow. 

We collect the edge TB cases in Table \ref{tab:edge} which shows that the LBL and the H-mode (ETB) have a low edge radial flux, i.e. a TB, whereas the TBL and the L-mode have a high edge radial flux.

\begin{table}[!ht]
\caption{Relationship between the fluid/plasma state and the edge radial flux.} 
\centering 
\begin{tabular}{llll} %
\hline\hline 
Edge state & Edge diffusion & Edge gradient & Edge radial flux \\  
& coefficient & & \\
\hline\hline 
FM: TBL & Large ($\nu_t$) & Steep & High \\
\hline
FM: LBL & Small ($\nu_{\rm kin}$) & Moderate & Low \\
\hline
PP: L-mode & Large & Moderate & High \\
\hline
PP: H-mode (ETB) & Small & Steep & Low \\
\hline 
\end{tabular}
\label{tab:edge} 
\end{table}

\subsection{Internal}

For FM, the wake in TBLs \cite{coles_a} can be modelled as a velocity jump or an internal shear layer \cite{krug_a}. For TWs, a step in the axial velocity profile has been observed, see e.g. Figure 21 (and 22) in \cite{wedin_a}. This is very similar to ITB profiles in PP and is a central observation of this paper. The wake can thus be interpreted as an ITB, possibly related to the quiescent core observed for channel flow \cite{kwon_a}.

We collect the core TB cases in Table \ref{tab:internal}, where we use the IIL term for FM. Diffusion coefficients are affected by RS driven flows and structures (TWs/magnetic islands) for both FM and PP. This is in contrast to the edge TBs, where the FM and PP mechanisms are different.

\begin{table}[!ht]
\caption{Relationship between the fluid/plasma state and the core radial flux.} 
\centering 
\begin{tabular}{llll} %
\hline\hline 
Core state & Core diffusion & Core gradient & Core radial flux\\  
& coefficient & & \\
\hline\hline 
FM: No IIL & Large & Moderate & High \\
\hline
FM: With IIL & Small & Steep & Low \\
\hline
PP: No ITB & Large & Moderate & High\\
\hline
PP: With ITB & Small & Steep & Low\\
\hline 
\end{tabular}
\label{tab:internal} 
\end{table}

\section{Core turbulence}
\label{sec:core}

In PP, turbulence during controlled confinement transitions has been studied by modifying the magnetic field structure \cite{zoletnik_a, basse_c}. It was found that core turbulence  increased for degraded confinement.

It has been speculated that a similar phenomenon occurs for the pipe flow high $Re$ transition \cite{basse_d}. Here, increased core turbulence for increasing $Re_{\tau}$ was observed, leading to the question whether low (high) $Re_{\tau}$ for pipe flow corresponds to good (bad) confinement in PP? Perhaps this points to the existence of a stronger (larger) wake for low $Re_{\tau}$.

\subsection{A possible re-interpretation of the high Reynolds number transition region}

The log-law region is associated with large turbulent structures and extends further inwards for higher $Re$. It can be argued that the high $Re$ transition corresponds to the log-law/wake transition region getting pushed towards the core or even collapses at sufficiently high $Re$. The loss of this IIL/ITB leads to higher levels of core turbulence, since the quiescent core is reduced or disappears and is replaced by structures from the log-law region. This is also reflected in an increasing turbulent length scale at the transition \cite{basse_d}. The proposed mechanism implies that turbulence in pipe flow ends up similar to open channel flow for sufficiently high $Re$ \cite{pirozzoli_a}. The transition from localised to expanding turbulence for the laminar-turbulent transition is associated with regions where the TKE production-to-dissipation ratio $\mathcal{P}/\varepsilon>1$ \cite{avila_a}, which corresponds to the non-equilibrium high $Re$ state detailed in \cite{basse_d}.

\section{An overview of concepts}
\label{sec:overview}

In this section we provide an appraisal of the similarities, differences and question marks we have identified so far.

\subsection{Similarities}

Possible FM/PP related flow phenomena are collected in Table \ref{tab:FM_PP_comp}. 

Regarding coherent (turbulent) structures, ECS and VLSM in FM can be considered counterparts of magnetic field structures in PP such as magnetic islands associated with rational surfaces or MHD phenomena.

Flow generation by RS has been observed as ZFs in PP and we argue that a similar mechanism is at play in FM.

As discussed, the IILs (momentum, heat, concentration) and ITBs have strong similarities. And the wake in FM can be likened to an ITB as proposed above. It is remarkable that the FM VF/TF model (Sections \ref{subsec:FM_UMZ} and \ref{subsec:FM_UTZ}) is almost identical to the q-comb model for PP (Section \ref{subsec:PP_safety}).

Finally, we have a separate table entry for RC structures observed for passive scalars and their similarity to sawtooth crashes in PP, which are magnetic reconnection events associated with particle and heat ejection from the plasma core.

\begin{table}[!ht]
\caption{Proposed analogies between FM and PP: Structures, flow and TBs.} 
\centering 
\begin{tabular}{ll} %
\hline\hline 
FM & PP \\  
\hline\hline
ECS, VLSM & Magnetic islands, MHD \\
\hline
RS driven flow & RS driven ZF \\
\hline
IIL & ITB \\
\hline
Wake & ITB \\
\hline
\hline
RC structures & Sawtooth crashes \\
\hline 
\end{tabular}
\label{tab:FM_PP_comp} 
\end{table}

The laminar-turbulent transition in FM can be correlated with confinement transitions in PP, see Table \ref{tab:FM_PP_conf}. Laminar flow is here interpreted as H-mode (low turbulence level) and turbulent flow as L-mode (high turbulence level). In PP, dithering between L- and H-mode can be thought of as collections of closely spaced ELMs \cite{basse_e,basse_f}, similar to puffs in FM. The relationship of increasing $Re$ in FM and worse confinement in PP is consistent with the findings presented in Section \ref{sec:core}.

\begin{table}[!ht]
\caption{Proposed analogies between FM and PP: Laminar-turbulent and confinement transitions.} 
\centering 
\begin{tabular}{lll} %
\hline\hline 
FM & PP & Edge radial \\  
& & flux \\
\hline\hline 
Laminar flow & H-mode & Low \\
\hline
Laminar regions & No ELMs & Low\\
Laminar-turbulent transition & H-L transition & \\
\hline
Turbulent puffs & ELMs & High\\
Laminar-turbulent transition & H-L transition & \\
\hline
Turbulent flow & L-mode & High \\
\hline 
\end{tabular}
\label{tab:FM_PP_conf} 
\end{table}

The mixing length concept from FM (Section \ref{subsec:FM_turb_len}) has been adopted in PP (Section \ref{subsec:PP_turb_len}). Apart from that it is difficult to compare important length scales; relevant micro-, meso- and macro-scales exist for both FM and PP, but other scales can also have an impact so a direct correspondence is not obvious.

Both FM and PP can be considered (quasi-) 2D, where the symmetry-breaking coordinate is streamwise for FM and toroidal (or parallel) for PP. Both FM and PP streamwise/toroidal turbulent structures can be extremely long, i.e. tens of pipe diameters for FM (VLSM) and tens of meters for PP. 

Cross-scale interaction between turbulent structures is important in both FM \cite{marusic_d} and PP \cite{ida_a,maeyama_a}.

\subsection{Differences}

Since we have compared pipe flow and toroidal devices, an obvious difference is the curvature. But we should mention that curved pipes and linear plasma machines exist and could be interesting to bring into this comparison.

The inboard/outboard asymmetry observed for PP, e.g. streamers, does not have a direct correspondence for FM. The closest would be the asymmetric TW identified in \cite{pringle_a}, where the TW only occupies half of the pipe.

Another important difference is the existence of EM fields for PP which are not present for FM. Thus, phenomena which can only be caused by EM fields are not relevant for our comparison. The associated twisted magnetic field lines in PP do not have an exact counterpart in FM; however, helical TWs exist with a similar structure \cite{pringle_a}.

The physical wall for FM, leading to the no-slip and no-penetration BCs, is non-existent at the separatrix for PP, where the conditions are free-slip and penetrable. The additional SOL for PP does not have an equivalent for FM; the only way to have fluid phenomena inside the wall is by use of image vorticity \cite{saffman_a} placed in nonphysical regions to satisfy the impermeable BC \cite{boatto_a} (pun intended).

\subsection{Question marks}

Particle mirror trapping in PP is caused by the magnetic field and there is no direct process like this in FM. The closest might be the rolls-streaks-waves SSP mentioned in Section \ref{subsec:FM_dynamical}.

Plasma shaping effects are important in PP, e.g. the GAM dependency on vertical plasma elongation. Many other shaping parameters exist, such as  triangularity, inverse aspect ratio and the Shafranov shift. Shaping also impacts FM flows, but this has not been explored in detail yet. 

The MFU in FM (Section \ref{subsec:FM_MFU}) may correspond to microscales in PP; here, additional investigations should be carried out (if they have not been so already?) to determine whether self-sustaining cycles exist in the plasma edge.

\section{Discussion}
\label{sec:disc}

Objectives in FM and PP can be different, but for both fields an understanding is needed no matter whether an effect has to be minimised or maximised.

Examples of quantities of importance for the two fields include:

\begin{itemize}
\item FM: Drag (pressure drop) and heat transfer
\item PP: Confinement: Cross-field anomalous transport of particles and heat
\end{itemize}

\subsection{Possible universal turbulent flow mechanisms}

We agree with the "chimera" view of turbulence proposed by Saffman \cite{saffman_b}
 \footnote{ \samepage "Finally, we should not altogether neglect the possibility that
there is no such thing as 'turbulence'. That is to say, it is not
meaningful to talk of the properties of a turbulent flow independently
of the physical situation in which it arises. In searching
for a theory of turbulence, perhaps we are looking for a chimera.
Turbulent phenomena of many types exist, and each one of practical
importance can be analysed or described to any required
degree of detail by the expenditure of sufficient effort. So perhaps
there is no 'real turbulence problem', but a large number
of turbulent flows and our problem is the self imposed and possibly
impossible task of fitting many phenomena into the Procrustean
bed of a universal turbulence theory. Individual flows
should then be treated on their merits and it should not necessarily
be assumed that ideas valid for one flow situation will
transfer to others. The turbulence problem may then be no
more than one of cataloguing, The evidence is against such an
extreme point of view as many universal features seem to exist, but nevertheless
cataloguing and classifying may be a more useful approach than we care to admit."} to some extent, but would argue that common ingredients do exist. These ingredients will change importance depending on the specific situation, but would always exist. If we continue using the chimera metaphor, this would be a chimera with the same body parts but different proportions. Traditional ingredients are:

\begin{itemize}
\item Geometry
\item BCs
\end{itemize}

And what has been argued in this paper are the additional (FM/PP) ingredients:

\begin{itemize}
\item ECS/Magnetic islands
\item RS driven (zonal) flows
\item IILs/ITBs
\end{itemize}

There is an interplay between the ingredients and some ingredients are less clear than others, for example whether the RS driven flow in FM and PP can be thought of as the same phenomenon. For the L-H transition in PP, "The picture is thus of an close interaction among
sheared flows, eddy structures, RS, and ZFs across the confinement transition" \cite{conway_a}.

\begin{figure}[!ht]
\centering
\includegraphics[width=12cm]{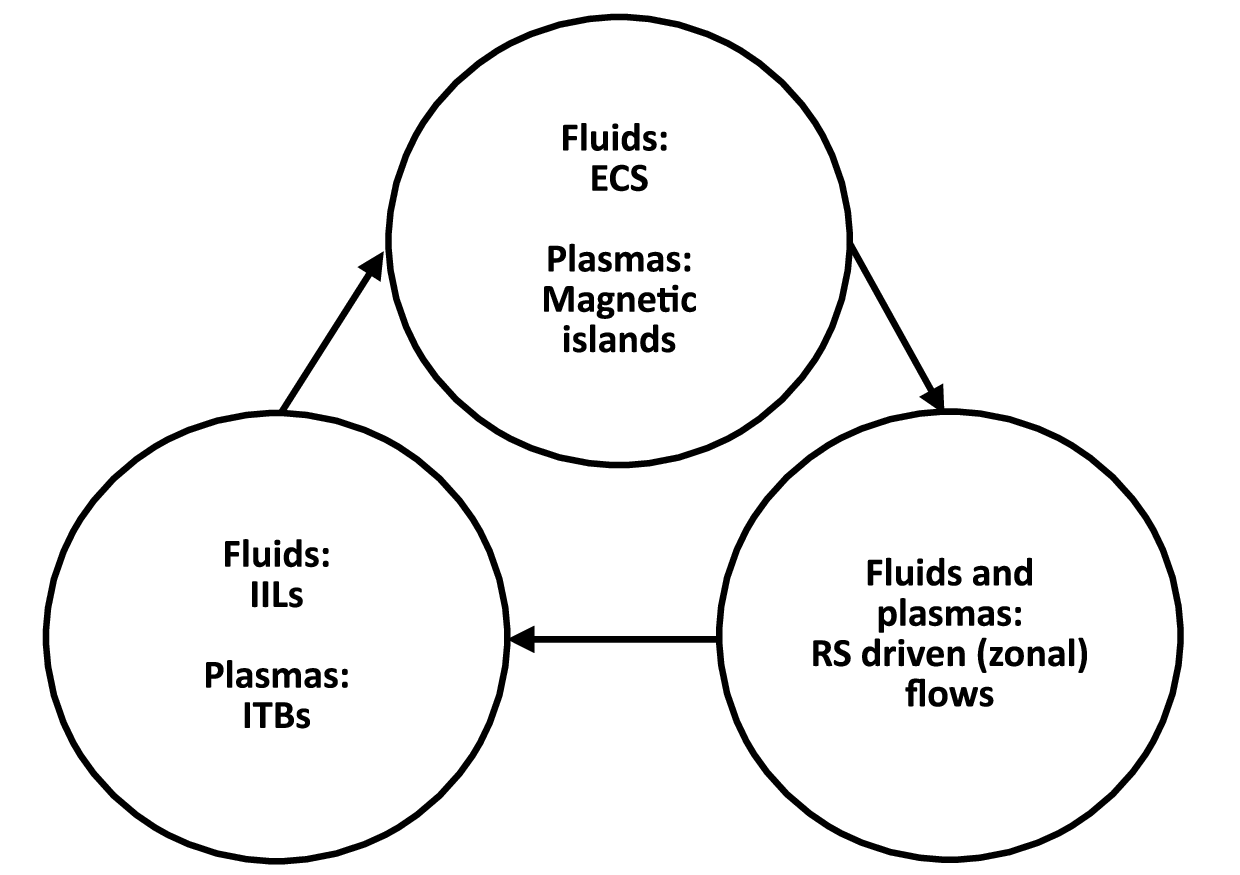}
\caption{Proposed common FM/PP process.}
\label{fig:proc}
\end{figure}

In Figure \ref{fig:proc}, we add a sketch which consists of the identified FM/PP ingredients. The arrows are meant to indicate the general sequence, but drawing such a picture opens up a wide range of new questions, which will be addressed in \cite{basse_g}.

\subsection{Nomenclature pertaining to radial fluxes}

There are different names to describe the manifestations of low or high radial fluxes in FM and PP, see Table \ref{tab:manifestations}. But the underlying mechanisms are the same, i.e. perpendicular/cross-field transport of momentum, particles and heat.

\begin{table}[!ht]
\caption{Manifestations of low and high radial fluxes.} 
\centering 
\begin{tabular}{lll} %
\hline\hline 
State & Radial flux & Manifestations\\  
\hline\hline 
FM: Laminar flow & Low & Small pressure drop\\
& & Low heat transfer\\
\hline
FM: Turbulent flow & High & Large pressure drop\\
& & High heat transfer\\
\hline
PP: Improved confinement & Low & Long energy and \\
(L-mode with ITB) & & particle confinement times\\
(H-mode with/without ITB) & & \\
\hline
PP: Degraded confinement & High & Short energy and \\
(L-mode without ITB) & & particle confinement times\\
\hline 
\end{tabular}
\label{tab:manifestations} 
\end{table}

\subsection{Proposal for new research program}

We recommend the initiation of specific cross-disciplinary efforts in e.g. model building for both the FM and PP communities to take advantage of progress in both fields.

An experimental FM approach would be to build a simple device to test e.g. ZF generation from RS without EM fields. This could either be a linear or toroidal device, where e.g. shape effects such as those found in PP could be tested using for example elliptical pipes.

In general, the cross-disciplinary work should cover all similarities, differences and question marks mentioned above, but focus on the question marks.

We end this section with three quotes on interesting avenues to take:

\begin{itemize}
\item \cite{terry_a}: "A simple, direct demonstration of shear suppression, ideally in a controlled neutral-fluid experiment, is a desirable direction for future work."
\item \cite{diamond_a}: "Finally, it must be said that the greatest opportunities for future research on zonal flows lie in the realm of experiment. Particular challenges include the simultaneous study, correlation and synthesis of generation dynamics in real space (i.e. via vorticity transport) and $k$-space (i.e. via nonlinear mode coupling), and the development of methods to control zonal flows. More generally, future experiments must emphasize challenging the theory and confronting it with stressful quantitative tests."
\item \cite{conway_a}: "A range of high quality diagnostics have been used in the study of ZFs, but, often lacking are comprehensive sets of simultaneous measurements of the flow oscillations, their structure (as well as their sidebands to confirm the ZFO or GAM identity), together with high-$k$ measurements of the ambient flow and density turbulence, its properties and structure." 
\begin{itemize}
(Here, ZFO are ZF oscillations.)
\end{itemize}
\end{itemize}

\section{Conclusions}
\label{sec:conc}

We have presented a comparative study of wall- and magnetically-bounded turbulent flows to identify possible cross-disciplinary similarities. The most important common phenomena found are coherent (turbulent) structures, shear Reynolds stress flow generation and transport barriers.

Exact coherent structures found in fluid mechanics appear to have many similarities with magnetic islands in fusion plasmas which are associated with rational values of the winding number of the magnetic field lines.

Zonal flows in fusion plasmas create radial velocity shear which is also seen between uniform momentum zones in nonionised turbulent flows. 

To the best of our knowledge, this is first time the uniform momentum zones in fluid mechanics have been compared to internal transport barriers in magnetically confined fusion plasmas.

We propose that these phenomena are common (universal) ingredients for both nonionised fluids and magnetically confined fusion plasmas:

\begin{itemize}
\item Exact coherent states/Magnetic islands
\item Shear Reynolds stress driven (zonal) flows
\item Internal interface layers (momentum, heat, concentration)/Internal transport barriers
\end{itemize}

The improved understanding has been used to re-interpret transport barriers and core turbulence. 

An additional potential similarity is between ramp-cliff structures in passive scalar flows and sawtooth crashes caused by magnetic reconnection in fusion plasmas.

Finally, we propose a new cross-disciplinary experimentally-based research program to test the ideas we have put forth.

{\it A note of caution:}
 Cross-disciplinary research is notoriously difficult both to carry out and to gauge, since you will be an outsider in some fields and risk being seen as a crackpot in others. This naturally leads to the disclaimer that all misunderstandings and errors are mine.

\paragraph{Acknowledgements}

Tak til C.V.J\o rgensen for "Alverdens turbulens".

\clearpage

\label{sec:refs}

\end{document}